\documentclass[journal]{IEEEtran}
\usepackage{cite}
\usepackage{textcomp}
\usepackage{amsmath,amssymb,amsfonts}
\usepackage{graphicx}
\usepackage{booktabs}
\usepackage{url}
\usepackage{xspace}
\usepackage{listings}
\usepackage{xcolor}
\usepackage{array}
\newcolumntype{L}[1]{>{\raggedright\arraybackslash}p{#1}}
\definecolor{lstcomment}{HTML}{6b6a66}
\lstdefinestyle{go}{
  language=Go, basicstyle=\scriptsize\ttfamily, columns=fullflexible,
  keepspaces=true, commentstyle=\color{lstcomment}\itshape,
  showstringspaces=false, breaklines=true, breakatwhitespace=true,
  frame=none, aboveskip=2pt, belowskip=2pt, xleftmargin=2pt}

\newcommand{\lieWindow}{196}
\newcommand{\corruptLo}{67} \newcommand{\corruptHi}{94}
\newcommand{\svFuseTrials}{7}
\newcommand{\svGrowthPerRestart}{18}
\newcommand{\svGrowthSeries}{node-disk usage from 10.0\,GB to 77.8\,GB over four restarts}
\newcommand{\svKubeletUsed}{741\,KB}
\newcommand{\svPartitionSize}{91.1\,GiB}
\newcommand{\svLazyCacheGiB}{14\,GiB}
\newcommand{\svLazyUsedGrowth}{2.3\,MB}
\newcommand{\svEvictedPods}{26}
\newcommand{\vanillaOkFiles}{7} \newcommand{\vanillaEIO}{273}
\newcommand{\oursOkFiles}{280}
\newcommand{\vanillaBytesRead}{4.1\,GB}
\newcommand{\oursBytesRead}{150.3\,GB}
\newcommand{\honestyPricePct}{18.5\%}
\newcommand{\honestyPriceLo}{15.3\%} \newcommand{\honestyPriceHi}{19.0\%}
\newcommand{\honestyPriceSinglePair}{20.6\%}
\newcommand{\sweepZeroPrefillS}{953.4} \newcommand{\sweepNinetyPrefillS}{1129.8}
\newcommand{\intPFiftyLo}{18.5\%} \newcommand{\intPFiftyHi}{18.8\%}
\newcommand{\intHitDropLo}{5.2} \newcommand{\intHitDropHi}{13.3}
\newcommand{\intReads}{4{,}100}
\newcommand{\thrashHotSetGB}{21.5\,GB} \newcommand{\thrashOccupiedGB}{73\,GB}
\newcommand{\thrashFactor}{3.4}
\newcommand{\ampLinearityLo}{$-$0.1\%} \newcommand{\ampLinearityHi}{+0.0\%}
\newcommand{\ampNicPct}{29\%}
\newcommand{\ampFasterLo}{1.1\%} \newcommand{\ampFasterHi}{5.6\%}
\newcommand{\writesSkippedLo}{4{,}079} \newcommand{\writesSkippedHi}{4{,}142}
\newcommand{\evictedUnderReadGB}{224.6\,GB}
\newcommand{\peakBudgetPct}{95.0\%}
\newcommand{\peakDuPct}{97.7\%}
\newcommand{\driftLo}{$-$7.75\%} \newcommand{\driftHi}{+0.08\%}
\newcommand{\cfourDiedDrops}{127{,}000}
\newcommand{\cfourHeldDrops}{301{,}000}
\newcommand{\cfourHeldDu}{76\,GB} \newcommand{\cfourBudget}{80\,GB}
\newcommand{\cfourResidualLo}{12.3\%} \newcommand{\cfourResidualHi}{12.8\%}
\newcommand{\policyGatePct}{4.7\%}
\newcommand{\policyGateFailPct}{6.1\%}
\newcommand{\abGapPct}{0.2\%}
\newcommand{\mbGetNs}{211} \newcommand{\mbGetPct}{1.9\%}
\newcommand{\mbAddUs}{1.28} \newcommand{\mbAddPct}{4.5\%}
\newcommand{\stressReads}{3.18\,M} \newcommand{\stressEvictions}{231{,}638}
\newcommand{\probeEIO}{272}

\newcommand{\estargz}{eStargz\xspace}
\newcommand{\stargzsn}{stargz-snapshotter\xspace}

\begin{document}

\title{Why Kubernetes Cannot Contain a Full Lazy-Pulling Cache: Two Impossibility Results and a Bounded, Honest Design for Model Serving}
\author{Georgii~Kliukovkin\thanks{G. Kliukovkin is an independent researcher in Sunnyvale, CA 94085 USA (e-mail: kliukovkin@gmail.com; ORCID 0009-0002-3323-1932).}\thanks{Preprint. This manuscript is under review at IEEE Access (manuscript Access-2026-57465, submitted 4 November 2026).}}

\markboth
{Kliukovkin: Why Kubernetes Cannot Contain a Full Lazy-Pulling Cache}
{Kliukovkin: Why Kubernetes Cannot Contain a Full Lazy-Pulling Cache}

\maketitle

\begin{abstract}
Lazy container-image pulling makes cold starts of autoscaled model-serving
fleets on Kubernetes nearly independent of model size, but it moves the
model's bytes into a node-level cache that the platform neither bounds nor
accounts for. This article establishes, by experiment, two impossibility
results for that cache. A monitoring-side remedy cannot catch its failure:
a sentinel readiness probe stays green while \probeEIO{} of 280 files are
unreadable. A platform-side remedy cannot contain it: kubelet, given full
disk-pressure visibility, frees 0 bytes by image garbage collection and
falls back to evicting pods (\svEvictedPods{} across the campaign, the
reading pod among them), because its reclamation actuator is keyed to an accounting
that does not know the cache exists. The article then eliminates the
failure class inside the component that owns the bytes, with no change to
kubelet or the container runtime interface, by two design rules: a byte
budget whose pressure signal is keyed to filesystem truth, so that a
never-blocking, lossy accounting still enforces a hard bound (held against
about \cfourHeldDrops{} dropped updates), and pass-through semantics under
which a failed cache write never becomes a failed read, with the coupling
traced to a single \texttt{io.MultiWriter}. Across six pre-specified
campaigns on AWS, upstream \stargzsn fails \vanillaEIO{} of 280 model files
at 90\% cache pre-fill; the proposed system reads all \oursBytesRead{} with
zero errors at +\honestyPricePct{} completion cost and holds the budget
under a live read, co-tenancy, and a three-node fleet. A source audit shows
the missing bound generalizes to SOCI and Nydus. The implementation is open
source and has been proposed upstream.
\end{abstract}


\section{Introduction}\label{sec:intro}
\IEEEPARstart{S}{erving} a large model on Kubernetes, on a managed cloud
service or a self-run cluster alike, begins with a delivery problem:
hundreds of gigabytes of weights must reach the node before the first
prediction.
Eager image pulling pays that cost up front and scales linearly with size;
lazy pulling---\estargz/\stargzsn~\cite{estargz} and AWS
SOCI~\cite{soci}---mounts the image immediately and fetches content on
demand, making cold time-to-first-prediction essentially independent of model
size~\cite{kliukovkin2026lazy}. For autoscaled and scale-to-zero serving
fleets, where every replica added under load pays the delivery cost before
serving a single request, this is the difference between seconds and tens of
minutes~\cite{kliukovkin2026coldstart}.

The price of that deferral is paid by a component most operators never
configure and the platform does not account for: the snapshotter's
node-level cache, shared by every lazily mounted image on the node, into
which every fetched chunk is written. In prior work we characterized what happens when
its volume fills under purely legitimate load~\cite{kliukovkin2026lazy}: pods
that are already \texttt{Ready} and serving begin failing reads of their
model files---\corruptLo--\corruptHi\% of files at high residual
occupancy---while the pod's readiness, health endpoint, and prediction
endpoint remain green (for at least \lieWindow{}\,s in our instrumented
timeline). The failure is invisible to every management-plane signal
Kubernetes exposes---readiness, liveness, prediction endpoints, node
conditions, events---and observable only in snapshotter-specific telemetry
that serving stacks do not collect by default. It is a fault with no
symptom at the layer that owns remediation.

That measurement study answered \emph{what happens}; this paper answers
\emph{what should}. It presents the design, implementation, and evaluation of
bounded, honest lazy pulling: a rework of the lazy cache's contract with the
platform above it, built as roughly 12{,}000 changed lines (about 40\%
tests) on \stargzsn and proposed upstream. No result below depends on the
under-review companion study: the vanilla arm of our evaluation
re-establishes the failure on this paper's own rigs
(Section~\ref{sec:eval}).

\textbf{Why not just monitor better, or let the platform react?} These
are the two remediation routes a cluster operator actually has, monitoring
and platform-driven eviction, and we close both empirically before
building. First, detection-side
mitigation is structurally limited: we built the natural mitigation---a
readiness probe that reads sentinel files through the mount---and it reported
green while \probeEIO{} of 280 files were unreadable, because damage
concentrates in file regions the probe's sample does not cover. A sampling
probe converts silent corruption into a dice roll, not a guarantee. Second,
platform-side remediation acts on the wrong accounting: with the cache
partition exported to kubelet's image filesystem accounting and production
eviction thresholds enabled, kubelet correctly raised
\texttt{NodeHasDiskPressure}---and then freed 0 bytes by image garbage
collection, because the cache is not image usage: the partition
(\svPartitionSize) was full while containerd reported \svKubeletUsed{} of
image usage; in a separate lazy-read run, about \svLazyCacheGiB{} of cache
accumulated while reported image usage grew by \svLazyUsedGrowth. kubelet
evicted \svEvictedPods{} pods across the runs (23 from one crash-evict
loop); none of the evictions could release any cache bytes. The platform
is \emph{sighted but powerless}: the gap is not signal; it is an actuator
keyed to accounting that does not know the cache exists. Elimination must
therefore happen inside the component that owns the bytes.

This paper makes four contributions:
\begin{itemize}
\item \textbf{Two measured impossibility results.} Monitoring-side and
  platform-side remediation, the two routes available to an
  operator, are each shown by experiment to be unable to
  contain this failure class: a sentinel probe stays green while
  \probeEIO/280 files are unreadable, and kubelet with full pressure
  visibility frees nothing by image garbage collection and falls back to
  evicting pods, none of which releases cache bytes. These generalize to any
  node-local cache outside the platform's reclamation accounting
  (Sections~\ref{sec:ledger},~\ref{sec:discussion}).
\item \textbf{A bounded cache that survives its own accounting.} Byte-budget
  eviction driven by a durable chunk index, with the pressure signal keyed
  to filesystem truth precisely because the index is lossy by contract: the
  filesystem decides \emph{how much}, the index decides \emph{which}. The
  budget held against $\sim$\cfourHeldDrops{} dropped accounting updates
  where a na\"ive rule died at $\sim$\cfourDiedDrops{}
  (Sections~\ref{sec:design},~\ref{sec:eval}).
\item \textbf{An honest data path, with its root cause.} Pass-through
  semantics---a chunk that cannot be cached is served to its caller and
  simply not kept---sever the failure structurally; the coupling they sever
  is traced to a single \texttt{io.MultiWriter} on the fetch path, with a
  regression test that fails on the parent commit and passes on our fix
  (Sections~\ref{sec:design},~\ref{sec:impl}).
\item \textbf{Evaluation on AWS against the failure suite of our
  companion study, through tenancy and fleet scale.} The before/after is run on identical rigs: upstream fails
  \vanillaEIO/280 files at 90\% pre-fill (three independent
  reproductions); our system serves 280/280 across the full
  \oursBytesRead{} read at +\honestyPricePct{} completion cost, with the
  budget never exceeded while \evictedUnderReadGB{} is evicted under the
  read. We report an honest negative on eviction-policy choice, and the
  operational findings the road here surfaced---including that the shipped
  systemd unit silently defeats the upstream FUSE manager
  (Section~\ref{sec:eval})---together with a source-level audit locating
  SOCI and both Nydus modes on the boundedness and coupling axes
  (Section~\ref{sec:audit}).
\end{itemize}

\section{Background and the Failure Ledger}\label{sec:ledger}

\subsection{Lazy pulling and the node-level cache}
An OCI image is a stack of compressed layer archives; an eager pull
downloads and unpacks all of them before the container starts. Lazy-pulling
snapshotters reformat (\estargz) or index (SOCI) the image so a FUSE
filesystem can be mounted immediately and individual chunks fetched from the
registry on first read. Every fetched chunk is written into a node-level
on-disk cache shared by all lazily mounted images on the node: for
\stargzsn, an HTTP blob-chunk cache and a filesystem cache under one root.
Model serving is the workload where both halves of this design are pushed
hardest: artifacts reach hundreds of gigabytes, autoscaling multiplies cold
starts, and a serving pod's correctness depends on every byte of its weights
being readable long after \texttt{Ready}.

\subsection{Five measured facts}\label{sec:facts}
The design in this paper is a response to five facts, each established by
controlled measurement; none is hypothetical. F1--F2 are from our prior
study~\cite{kliukovkin2026lazy}; F3--F5 from the pre-design measurement
campaign for this work (rig details in Section~\ref{sec:eval}).
Table~\ref{tab:ledger} summarizes the ledger and the requirement each fact
imposes on the design.

\begin{table}[t]
\caption{The failure ledger: each measured fact and the design requirement
it imposes. F1--F2 from~\cite{kliukovkin2026lazy}; F3--F5 from this work's
pre-design campaign.}
\centering
\setlength{\tabcolsep}{3.5pt}
\scriptsize
\begin{tabular}{@{}l L{2.6cm} L{1.9cm} L{2.4cm}@{}}
\toprule
 & Measured fact & Key evidence & Requirement \\
\midrule
F1 & cache unbounded on disk & 74 config keys, none bound bytes &
bound it at all \\
F2 & exhaustion corrupts \emph{already-serving} pods &
\corruptLo--\corruptHi\% files fail; signals green $\geq$\lieWindow\,s &
decouple cache failure from read failure \\
F3 & sampling probes protect their sentinels, not the model &
probe green at \probeEIO/280 unreadable &
decouple by construction, not monitoring \\
F4 & platform sighted but powerless &
kubelet freed 0\,B; \svEvictedPods{} evictions released no cache bytes &
bound must live where the bytes live \\
F5 & shipped remediation defeats itself &
\svFuseTrials/\svFuseTrials{} KillMode trials; up to
+\svGrowthPerRestart\,GB per graceful restart &
bound must survive restarts \\
\bottomrule
\end{tabular}
\label{tab:ledger}
\end{table}

\textbf{F1: The cache is unbounded on disk.} No configuration key bounds
the on-disk size of \stargzsn's cache tree: the two cache-sizing knobs bound
in-memory LRU entries and open file descriptors. We verified the key surface
of the development head is byte-identical to the release we measured---74
keys, none size-related. Cache directories are cleaned only per-layer when
an image is removed.

\textbf{F2: Exhaustion corrupts the already-serving.} When the cache volume
fills under sustained legitimate reads, running pods fail reads of model
files---\corruptLo--\corruptHi\% of files, scaling monotonically with
residual occupancy---while readiness, health, and prediction endpoints stay
green (\lieWindow{}\,s lower bound in an instrumented timeline). Damage
concentrates in file \emph{tails}: after one induction, every one of 280
files passed a 1\,MiB read at offset zero while 262 failed a full read. Any
health check that samples is therefore structurally biased toward the
cached head of the data.

\textbf{F3: Detection-side mitigation fails by sampling.} We implemented the
natural fix---a sentinel readiness probe that tail-reads a fixed set of
files through the mount---and pre-specified its falsifier. It reported
green while \probeEIO{} of 280 files were unreadable: the probe protects its
sentinels, not the model. This is F2's bias, weaponized against the
monitor itself.

\textbf{F4: The platform is sighted but powerless.} We exported the cache
partition to kubelet's image-filesystem accounting (the mechanism upstream
added for disk-space visibility) and enabled production eviction thresholds
(\texttt{imagefs.available} 10\%), then drove the partition past the
threshold in two runs. In the first, a lazy read on top of a 75\%
\texttt{fallocate} filler grew the cache until kubelet evicted the reading
pod. In the second, after a reset, a \texttt{fallocate} filler took the
partition to 100\%, so that the reclamation chain could be observed
deterministically. The full chain fired:
\texttt{NodeHasDiskPressure} was raised; image garbage collection ran---and
freed \textbf{0 bytes}, because the snapshotter's content cache is not
``images'' in containerd's accounting: the partition (\svPartitionSize) was
full while containerd reported \svKubeletUsed{} of image usage; in the
separate lazy-read run, about \svLazyCacheGiB{} of cache accumulated while
reported image usage grew by \svLazyUsedGrowth. kubelet then fell through
to its only remaining actuator: it evicted \svEvictedPods{} pods across the
runs (23 from one crash-evict loop); none of the evictions could release
any cache bytes, while a fresh lazy pod stayed
\texttt{Pending} indefinitely. The gap is not signal; it is an actuator
keyed to the wrong accounting. Remediation must live in the component that
owns the bytes.

\textbf{F5: The shipped remediation defeats itself.} \stargzsn ships an
opt-in FUSE manager whose purpose is to keep mounts alive across daemon
restarts. In \svFuseTrials/\svFuseTrials{} pre-specified trials, enabling
it under the project's own systemd unit changed nothing: the unit's default
control-group kill mode SIGTERMs the detached manager on every
\texttt{systemctl restart}, and its SIGTERM handler unmounts everything
(the manager escapes the process group, not the cgroup). The consequence
is measured, not inferred: in the control arm every restart left running
pods with dead mounts---8/8 model files failing with \texttt{ESTALE} from
the first restart on. Under a corrected
unit the manager works---and then each \emph{graceful} restart duplicates
cache contents in place, up to \svGrowthPerRestart\,GB per steady-state
restart with zero application reads between restarts (\svGrowthSeries),
where upstream documentation acknowledges duplication only for unexpected
termination.
Both findings are reported upstream with fixes.

Together the ledger defines the requirement: the cache must be bounded by
the component that owns it (F1, F4), failure of the cache must be decoupled
from failure of reads (F2, F3), and the bound must survive restarts and the
accounting's own imperfections (F5).

\subsection{Is the class shared? A source audit of SOCI and Nydus}
\label{sec:audit}
The ledger is measured on one snapshotter. To locate the rest of the
ecosystem in it, we audited soci-snapshotter and Nydus (nydusd plus
nydus-snapshotter, in both its RAFS-userspace and EROFS-over-fscache
modes) at pinned releases---statically, source only, every claim carrying
a file-and-line citation in the released audit document. The result
(Table~\ref{tab:audit}) is not a uniform ``they all share it,'' and is
more useful for that.

\textbf{F1 generalizes.} Neither SOCI (72 leaf configuration keys; the
LRU knobs bound in-memory entries and descriptors, and LRU eviction
never unlinks a cache file) nor Nydus bounds the on-disk bytes of its
cache; the one partial exception, fscache mode, delegates the bound to
the kernel's \texttt{cachefiles} culling---whole-filesystem watermarks,
configured outside either project, with the snapshotter unable even to
measure usage. Sharper than the missing knob: both projects ship a
\emph{dead} cache-management key---SOCI's \texttt{http\_cache\_type} is
read by nothing, and nydus-snapshotter's \texttt{cache\_manager.gc\_period}
is stored and never read (no ticker exists)---so the configuration
surface advertises management that is not implemented.

\textbf{F2 is a design axis, not a universal.} The shared property is
\emph{placement}: whether the cache write sits on the delivery path.
SOCI couples more strongly than stargz---the span is fetched in full
into memory and then \emph{discarded} when its cache write fails,
surfacing as \texttt{EIO} for bytes that had arrived. Nydus's RAFS mode
sits on the other end: the reader is filled from the fetched buffer
while the cache write is dispatched asynchronously, so a persist failure
costs a cache entry, never a read---a production existence proof that
the decoupling this paper argues for is deployable (one deprecated
option, \texttt{cache.compressed}, re-couples it). Its fscache mode
makes the failure invisible instead: a populate error is logged and the
kernel is acknowledged unconditionally; what the reader then observes is
decided kernel-side and is not determinable statically.

\textbf{F5 splits along a one-sentence design lesson.} The two systems
that duplicate cache across restarts (stargz, SOCI---the latter
unconditionally: a fresh temporary directory per resolution, no startup
sweep, and \texttt{SIGTERM} skips cleanup entirely) are exactly the two
that name cache directories non-deterministically; the two that re-adopt
(both Nydus modes) derive the cache path from the content identity.

\begin{table*}[t]
\caption{Source audit at pinned releases: does the rest of the
lazy-pulling ecosystem share the failure ledger? Static analysis; every
cell is backed by file:line citations in the released audit document.}
\centering
\setlength{\tabcolsep}{4pt}
\scriptsize
\begin{tabular}{@{}l L{4.2cm} L{5.2cm} L{4.4cm}@{}}
\toprule
System (pinned) & On-disk byte bound? & Cache-write failure can fail a
read? & Restart re-adoption? \\
\midrule
stargz v0.18.2 & no: 74 keys, none size-related (F1) &
yes: writer coupling on the delivery path (F2) &
no: duplicates in place, up to \svGrowthPerRestart\,GB per steady-state restart (F5) \\
SOCI v0.15.0 & no: 72 leaf keys; LRU knobs bound in-memory entries and
FDs; LRU eviction never unlinks a file &
yes, more strongly: the fully fetched span is discarded when its cache
write fails; surfaces as \texttt{EIO} &
no, unconditionally: fresh temp dir per resolution, no startup sweep,
\texttt{SIGTERM} skips cleanup \\
Nydus RAFS v2.4.5 & no: no byte key; snapshotter \texttt{gc\_period} is
dead code &
no (default): reader served from the fetched buffer, persist is
asynchronous; deprecated \texttt{cache.compressed} re-couples &
yes: content-derived path, persisted chunk map \\
Nydus fscache v2.4.5 & external only: kernel \texttt{cachefiles}
culling, outside the snapshotter's control or visibility &
invisible: populate errors logged, kernel acknowledged unconditionally;
reader-visible effect is kernel-side (not determinable statically) &
yes: readiness rebuilt from the backing file's extents \\
\bottomrule
\end{tabular}
\label{tab:audit}
\end{table*}

\section{Design}\label{sec:design}
Three principles organize the system, implemented as roughly 12{,}000
changed lines on \stargzsn (about 40\% tests); all behavior is opt-in and
off by default.

\subsection{Durable, non-blocking accounting}
A persistent index (bbolt) records every cached chunk---path, size, last
observed access, in-flight pin---maintained write-through from the cache's
data path via a bounded queue that \emph{never blocks and never fails a
cache operation}: updates are dropped, counted, when the queue is full.
Recency lives in the index, not filesystem \texttt{atime} (fleet nodes mount
\texttt{noatime}). On startup the index is validated (including a manual
check of bbolt's meta pages, since a truncated database faults the process
rather than erroring) and rebuilt by a background scan when missing or
inconsistent---which is also how a restarted daemon re-adopts existing
cache instead of accounting beside it (F5's growth).

Two consequences of ``never block'' are worth naming because the
evaluation later turns on them. First, the queue is bounded (65{,}536
entries) and sheds load by dropping, so the index is \emph{lossy by
contract}: correctness of enforcement cannot be allowed to depend on its
completeness (Section~\ref{sec:c4} tests exactly this). Second, because
updates are asynchronous, the index may momentarily trail the
filesystem in either direction; every consumer of the index is written
against ``advisory'' semantics, and the invariants below are chosen so
that staleness only ever costs a missed eviction candidate or a
redundant refetch, never a wrong byte. The accounting costs
+\mbGetNs\,ns (+\mbGetPct) on the cache-hit read path and
+\mbAddUs\,\textmu s (+\mbAddPct) on the write path in microbenchmarks.

\subsection{Byte-budget eviction, keyed to filesystem truth}
Eviction runs inside the index's single writer goroutine---no new locks on
the data path. When pressure exceeds a high watermark, victims are selected
by a pluggable policy over the index snapshot (pinned entries excluded) and
removed until a low watermark holds; unlinking during concurrent reads is
safe because POSIX keeps data alive under open descriptors, and a
subsequently needed chunk simply refetches through the existing miss path.

Six invariants, fixed at design time and each carried by a specific
mechanism rather than by care, define eviction's safety envelope:

\begin{itemize}
\item[\textbf{I1}] \textbf{Never evict a pinned chunk.} A pin covers the
  in-flight window between staging-file creation and rename-commit; the
  index is the only victim source and pinned entries are excluded from
  policy iteration.
\item[\textbf{I2}] \textbf{Eviction cannot race commit.} A chunk enters
  the index only after its rename-commit, so a path not yet committed
  cannot be selected. The stale direction---committed but not yet applied
  to the index---only makes eviction \emph{miss} a candidate, which is
  safe.
\item[\textbf{I3}] \textbf{Unlink during read is safe.} Reads go through
  an open descriptor, and POSIX keeps unlinked data alive until last
  close; eviction therefore takes no read-side locks at all.
\item[\textbf{I4}] \textbf{Refetch is the existing miss path.} The next
  read of an evicted chunk sees \texttt{ENOENT} and falls into the
  ordinary cache-miss branch; eviction adds no new code to the hot path,
  and its cost is a refetch, which the evaluation prices.
\item[\textbf{I5}] \textbf{Crash safety is inherited from the advisory
  index.} A crash between unlink and index-delete leaves a dangling
  entry; the next read corrects it as a miss and the rebuild scan
  reconciles fully. No ordering constraints, no \texttt{fsync} on the
  eviction path.
\item[\textbf{I6}] \textbf{Self-protection is policy, not mechanism.} A
  140\,GB sweep through an 80\,GB budget must evict its own tail; whether
  it may also flush other images' resident hot sets is the eviction
  policy's business. The mechanism guarantees only watermark
  convergence.
\end{itemize}

The subtle question is what ``pressure'' is. The index under-counts by
design whenever updates are dropped, and an eviction engine acting on an
under-count re-creates F2 with extra steps: it stops evicting while the
partition fills. Our first implementation had exactly this defect; a
measurement campaign found it (Section~\ref{sec:eval}). The corrected rule:
pressure is the index total plus a conservative price for every dropped
update, raised further to the filesystem's own measurement of the cache's
usage whenever the index is known to be lossy---and when eviction has
surrendered every chunk it knows about while the filesystem still reports
over-budget, the index itself is declared wrong and a rate-limited rescan
rebuilds it, with eviction re-run against the result. The filesystem is
authoritative for \emph{how much}; the index only ever decides
\emph{which}.

\subsection{Honesty: cache failure is never read failure}
The load-bearing principle. On the fetch path, bytes stream from the
registry simultaneously into the cache writer and the caller's buffer; in
upstream code one \texttt{io.MultiWriter} couples them, so a cache-side
write error stops delivery to a caller who needed nothing from the cache
(Section~\ref{sec:impl}). Our pass-through semantics: a cache write that
fails---after one emergency eviction cycle and one retry at the repeatable
points---fails \emph{only the caching}. The fetched bytes are served, the
chunk is simply not kept, a counter records the skip, and the very next
chunk is offered to the cache again. Exhaustion becomes a latency
degradation with a metric, never a correctness event. The residual honesty
surface is deliberately modest: budget occupancy, evictions, skipped
writes, and per-\texttt{errno} fetch failures as Prometheus series, plus a
node-events agent; given F3, we explicitly do \emph{not} offer a probe as a
safety mechanism.

\section{Implementation}\label{sec:impl}

\subsection{The root cause, and severing it}\label{sec:rootcause}
Upstream's fetch path streams a range response into the cache and the
caller simultaneously: for each chunk, a cache writer is opened and the
response body is copied through
\texttt{io.MultiWriter(cacheWriter, callerBuffer)}.
\texttt{io.MultiWriter} stops at the first writer that returns an
error---so an \texttt{ENOSPC} raised by the cache at any of its three
failure points (opening the staging file, the copy itself, or the commit
rename) stops the same bytes from reaching a caller that needed nothing
from the cache. In the commit case the caller's buffer is already complete
and is discarded anyway. The error propagates through the FUSE layer as
\texttt{EIO} on a read of a file whose bytes had, in fact, arrived. This is
the entire mechanism behind F2: a full cache does not merely stop caching;
it vetoes reads.

Fig.~\ref{lst:sever} shows the coupling and its severing, condensed
from the actual patch. The fix wraps the cache writer in a \emph{tolerant
writer} that remembers its first failure and reports full progress, so the
caller's copy always completes; the staging file is then aborted---never
committed partially---and a per-\texttt{errno} counter records the skip.

\begin{figure}[t]
\begin{lstlisting}[style=go]
// UPSTREAM: one writer chain couples cache and caller.
w := io.Writer(cw)             // cache writer
if _, ok := fetched[chunk]; ok {
    w = io.MultiWriter(w, allData[chunk])
}
if _, err := io.CopyN(w, r, chunk.size()); err != nil {
    cw.Abort()
    return err   // ENOSPC in the cache fails the READ
}

// OURS: the cache may fail; the caller's copy cannot.
type tolerantWriter struct { w io.Writer; err error }

func (t *tolerantWriter) Write(p []byte) (int, error) {
    if t.err != nil { return len(p), nil }
    if _, err := t.w.Write(p); err != nil { t.err = err }
    return len(p), nil       // always claims full progress
}

cached := &tolerantWriter{w: cw}
var w io.Writer = cached
if dest != nil { w = io.MultiWriter(cached, dest) }
if _, err := io.CopyN(w, r, chunk.size()); err != nil {
    cw.Abort()
    return err   // the FETCH failed: nothing to serve
}
if cached.err != nil {       // cache failed, caller is whole
    cw.Abort()
    return b.skippedCaching(chunk, fetched, cached.err)
}
\end{lstlisting}
\caption{The root cause and its severing (condensed from the patch to the
chunk-caching path). Upstream, \texttt{io.MultiWriter} stops at the first
failing writer, so a cache-side \texttt{ENOSPC} halts delivery to the
caller mid-copy. With the tolerant wrapper, only a failure of the fetch
itself can fail the read.}
\label{lst:sever}
\end{figure} Chunks that
arrive on the wire but have no live caller are drained to keep the
multi-chunk response body in sync. A caller that joined another caller's
in-flight fetch and finds the result uncached retries a bounded number of
times and then fetches its range directly: a read is never failed because
the cache could not keep something. A regression test asserting
``\texttt{ENOSPC} in the cache, bytes still served'' fails on the parent
commit with the original error string and passes on the fix.

The property that generalizes is placement, not the particular API.
SOCI couples the same way with no \texttt{io.MultiWriter} at all: the
fully fetched span is discarded when its cache write fails
(Section~\ref{sec:audit}). Nydus's RAFS mode shows the opposite
placement shipping in production---reader served from the fetched
buffer, persist asynchronous. The design question for any lazy-pulling
cache is therefore not ``does it use \texttt{MultiWriter}'' but ``is the
cache write on the delivery path.''

\subsection{Hardening the restart path}
Two defects found by our own measurement campaigns shaped the final
implementation. First, the index handover on daemon reconnect originally
opened the new index before releasing the old one; the newcomer could lose
the database lock, silently leaving the node with a configured budget and
no eviction. The fix orders release before construction and adds a bounded
retry, with two deliberately distinct alarm series: a retry that
\emph{succeeded} means handover is racing (a bug to report), a retry that
gave up means the node is running unenforced (an outage). Second, the
pressure calculation of Section~\ref{sec:design} exists because its
na\"ive predecessor---trusting the index alone---was demonstrated to let
the budget fail exactly when protection matters most
(Section~\ref{sec:c4}).

\subsection{Observability surface}
The honesty principle demotes monitoring from safety mechanism to
diagnosis, and the shipped surface reflects that. Prometheus series cover
budget occupancy and configured budget, eviction counts and bytes by
policy, skipped cache writes, dropped accounting updates, rescan runs,
and blob-fetch failures by \texttt{errno}; a small node-events agent
mirrors threshold crossings (budget exceeded, enforcement unavailable,
rescan escalation) into Kubernetes events so they surface where operators
already look. The sentinel-file probe of F3 ships only as an explicitly
labeled smoke check: after its falsification we do not offer any sampling
probe as a correctness signal, and the two handover-alarm series of the
previous subsection distinguish ``racing but recovered'' from ``running
unenforced''---the alert that matters most, because an unenforced node
looks healthy until F2.

\subsection{Upstream status}
The work is public as a branch on a fork of \stargzsn. The pass-through
fix of Section~\ref{sec:rootcause} and the first observability slice
(per-\texttt{errno} blob-fetch failure counter) are open upstream pull
requests; the byte bound is proposed upstream as a design issue; the two
operational findings of F5 are filed upstream with data and one-line
fixes; and the SOCI coupling finding of Section~\ref{sec:audit} is filed
with the SOCI project with file-and-line evidence. Nothing in this
paper's evaluation depends on merge status.

\section{Evaluation}\label{sec:eval}

\subsection{Method}
All experiments run on a two-host rig in one AWS availability zone
(us-east-1): a registry host and a node host, both i4i.2xlarge (8 vCPU,
64\,GB RAM, local NVMe), kind/Kubernetes v1.36, containerd 2.3.1, with the
snapshotter's cache on a dedicated 92\,GB NVMe partition and a configured
budget of 80\,GB (watermarks 0.95/0.85). The dedicated registry host
keeps fetch latency constant across arms, and the dedicated cache
partition means exhaustion experiments cannot side-load the OS volume.
The workload is the failure suite of our companion
study~\cite{kliukovkin2026lazy}: a
140\,GB-class \estargz model image of 280 files with a data-path-coupled
predictor, so a read error is observed as a serving failure, not
inferred from logs. The rig uses kind on EC2 rather than a managed Kubernetes service; the
F4 and elimination results do not depend on that choice. The mismatch F4
measures is structural (the snapshotter's cache tree is not part of
containerd's image store on any node type, and kubelet's split image
filesystem (KEP-4191) accounts for containerd's image store, not for the
snapshotter's cache beside it), the kubelet eviction chain exercised is
the same code a production node runs, and the elimination mechanism
lives entirely in the snapshotter, below the control plane.

Expectations for every campaign were pre-specified: written protocols
with pass and fail criteria, fixed before the rig for that campaign
existed and released with the artifact. (We say pre-specified rather
than pre-registered: no third-party registry was used.) Each trial
leaves an evidence bundle (journals, metric
snapshots, \texttt{df}/\texttt{du}/\texttt{lsof} samples, probe
transcripts) collected immediately at trial end, a rule instituted after
a watchdog teardown cost one campaign its artifacts. Invalidated trials
are preserved and flagged, not deleted; all raw data, evidence bundles,
and the harness are released. One methodological rule
was itself measured into existence: identical builds differed by up to
16\% in absolute completion time across rig instances while an interleaved
same-session A/B of two different builds differed by \abGapPct---so every
comparison below is within-session, and no cross-instance absolute is
quoted as a comparison.

The results below come from six campaigns run in sequence
(Table~\ref{tab:campaigns}). Two of them falsified our own
implementation---the na\"ive pressure rule and a lock-handover race---and
the fixes were then confirmed under re-runs of the same pre-specified
protocol. We report the campaign structure because the defects it caught
are themselves part of the design argument
(Sections~\ref{sec:design},~\ref{sec:c4}).

\begin{table}[t]
\caption{The six evaluation campaigns. Each ran against pre-specified
expectations; two falsified our own implementation, and the fixes were
re-confirmed under the same protocol.}
\centering
\setlength{\tabcolsep}{3.5pt}
\scriptsize
\begin{tabular}{@{}l L{3.1cm} L{3.6cm}@{}}
\toprule
 & Pre-specified question & Outcome \\
\midrule
I & Can the platform (kubelet) or the shipped FUSE manager already contain
the failure? & No: F4 and F5 established
(Section~\ref{sec:facts}). \\
II & Does pass-through eliminate the failure class? Does a sentinel probe
suffice? & Elimination confirmed (Table~\ref{tab:elim}); probe falsified
(F3); live-budget sweep held (Fig.~\ref{fig:sweep}); found an index
lock-handover race. \\
III & Does the budget survive its own lossy accounting? & No---na\"ive
pressure rule died at $\sim$\cfourDiedDrops{} drops
(Section~\ref{sec:c4}); handover race root-caused. \\
IV & Do the fixes hold under the original protocols? & Corrected pressure
held at $\sim$\cfourHeldDrops{} drops; handover clean (0 retries);
restart growth flattened (Fig.~\ref{fig:restart}); policy arms gated at
\policyGatePct. \\
V & Does a contended regime separate LRU from 2Q? & Gate failed (\policyGateFailPct);
reported only as consistent with the null. \\
VI & Does the bound survive a second tenant and fleet-level concurrency?
Is registry egress linear in node count? & Zero read errors under both;
interference modest and replicated (Section~\ref{sec:tenancy}); egress
linear to within 0.1\%; the slowdown clause falsified ($k{=}3$ ran
\emph{faster}, cause uninstrumented and reported as such); found:
a single tenant's hot set occupies \thrashFactor$\times$ its bytes. \\
\bottomrule
\end{tabular}
\label{tab:campaigns}
\end{table}

\subsection{Elimination of the failure class}
Table~\ref{tab:elim} is the headline. At 90\% cache pre-fill, upstream
v0.18.2 reads \vanillaBytesRead{} before the sweep collapses and fails
\vanillaEIO{} of 280 model files with \texttt{EIO}---a result reproduced
with identical counts on three independent rig instances across three
campaigns. Identical counts across independent instances are strong
evidence against a timing race on this configuration and consistent with
a structural mechanism---which files survive appears fixed by the
geometry of chunk boundaries against the pre-fill (we compared counts,
not file identities, so set-level determinism is inferred, not
verified): identically provisioned nodes under this workload fail
identically. Our system, on the same rig in the same session, reads the
full \oursBytesRead{} with \textbf{zero} errors, in both replicates. The
\writesSkippedLo--\writesSkippedHi{} skipped cache writes per run confirm
the out-of-space path was genuinely exercised---the zero-error result is
pass-through working, not pressure failing to materialize. The price of
honesty was replicated as five interleaved within-session pairs of the
same full read at 0\% versus 90\% pre-fill (medians
\sweepZeroPrefillS\,s versus \sweepNinetyPrefillS\,s): median
\textbf{+\honestyPricePct}, range +\honestyPriceLo{} to
+\honestyPriceHi, all ten runs 280/280 with zero errors. An earlier
single pair on a different instance had measured
+\honestyPriceSinglePair---outside the replicated range---so the ratio
itself varies by instance, and we quote only the replicated median.
Exhaustion has become a latency tax priced by registry proximity: on this
rig's same-AZ registry it is +\honestyPricePct; against a distant
registry every refetch pays the longer round trip, a cost we do not
measure here.

\begin{table}[t]
\caption{Elimination result: full read of a 140\,GB-class model (280
files) at 90\% cache pre-fill, within-session pairs, N=2 each. The
vanilla row is the within-session pair member for the quoted runs; the
two further vanilla reproductions on other instances (identical counts)
are corroboration, not comparison anchors.}
\centering
\setlength{\tabcolsep}{4pt}
\begin{tabular}{lcccc}
\toprule
 & files OK & I/O errors & bytes read & writes skipped \\
\midrule
vanilla v0.18.2 & \vanillaOkFiles{} / 280 & \vanillaEIO & \vanillaBytesRead & 0 \\
ours & \textbf{\oursOkFiles{} / 280} & \textbf{0} & \oursBytesRead & \writesSkippedLo--\writesSkippedHi \\
\bottomrule
\end{tabular}
\label{tab:elim}
\end{table}

\subsection{Bounded at full scale}
\begin{figure}[t]
\centering
\includegraphics[width=\columnwidth]{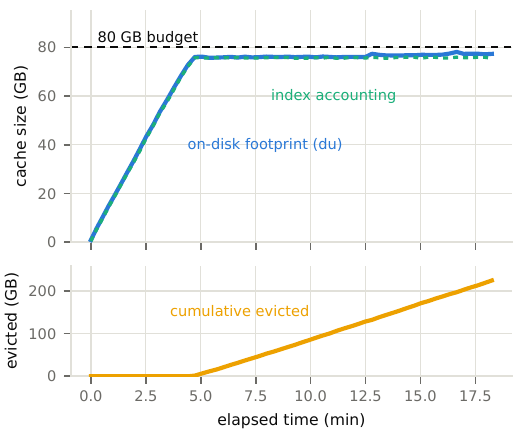}
\caption{An 80\,GB budget under a live 140\,GB sweep (59 samples over one
full-model read). Top: on-disk cache footprint (\texttt{du}) and the
index's own accounting converge just under the budget at the high
watermark; index-reported peak \peakBudgetPct{} of budget, on-disk peak
\peakDuPct. Bottom: \evictedUnderReadGB{} is evicted beneath the
read---with zero read errors.}
\label{fig:sweep}
\end{figure}

Fig.~\ref{fig:sweep} shows the budget under the full 140\,GB sweep with an
80\,GB budget: the engine evicted \evictedUnderReadGB{} beneath the live
read---every byte of it refetchable, none of it vetoing a read---while
index-reported occupancy peaked at \peakBudgetPct{} of budget (pinned at
the high watermark) and the on-disk footprint peaked at \peakDuPct,
never crossing the budget (59 samples; accounting drift versus
\texttt{du} between \driftLo{} and \driftHi{} excluding the first,
near-empty sample, where the asynchronous index trails a cold cache's
opening seconds by design). Two transient blob-fetch errors early in the
run were absorbed by the retry path and never surfaced: all 280 files
read cleanly. The two occupancy curves in the figure are the design's
split made visible: the index's accounting tracks the filesystem's own
measurement closely under healthy operation, and the pressure rule of
Section~\ref{sec:design} exists for exactly the moments they diverge.

\begin{figure}[t]
\centering
\includegraphics[width=\columnwidth]{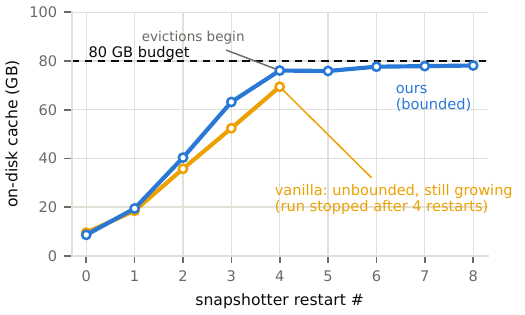}
\caption{Restart-driven cache duplication (F5), vanilla versus ours.
The y-axis is the cache tree's on-disk footprint (\texttt{du}); the
vanilla node's partition itself stood at 84\% used when its run was
stopped after four restarts. Vanilla with a working FUSE manager
duplicates cache contents on every \emph{graceful} restart with zero
application reads in between; our re-adoption scan plus byte budget
flattens the same pathology just under the budget, with zero read errors
across eight restarts.}
\label{fig:restart}
\end{figure}

The restart pathology of F5 is capped by the same mechanism
(Fig.~\ref{fig:restart}): where the working FUSE manager grew the cache
by up to \svGrowthPerRestart\,GB per graceful restart, unbounded
(\svGrowthSeries---84\% of its cache partition and still climbing when
the run was stopped), our build's re-adoption scan plus budget flattens occupancy
just under the budget from the fourth restart on, with zero read errors
across eight restarts. A hypothesis that the duplication was
deleted-but-open inodes was tested and refuted (\texttt{lsof} accounting
showed 0 bytes at every step); the growth is consistent with duplicate
chunk data on disk, which is precisely what a byte budget can reclaim
regardless of its exact composition.

\subsection{Surviving lossy accounting}\label{sec:c4}
The design's hardest test was self-inflicted. The accounting queue drops
updates rather than block the data path; our first eviction engine trusted
the resulting under-count, and a write-flood campaign killed it: at
$\sim$\cfourDiedDrops{} dropped updates the partition reached 100\% and the
snapshotter could no longer start. The corrected pressure rule
(Section~\ref{sec:design}) was then driven with $\sim$\cfourHeldDrops{}
dropped updates through the same flood: \texttt{du} held at \cfourHeldDu{}
against the \cfourBudget{} budget, the node remained startable, and the
escalation rescan fired 3--4 times per arm during enforcement (no arm
ran with the rescan disabled, so its necessity is inferred from the
design, not ablated). Table~\ref{tab:cfour} contrasts the two rules
under the same flood protocol; the arms ran in successive campaigns, but
the comparison is categorical (fatal versus held) and the corrected rule
absorbed $2.4\times$ the drops that killed its predecessor, so
cross-session placement cannot manufacture the result.

\begin{table}[t]
\caption{Write-flood against the budget: the same protocol, before and
after the pressure rule was keyed to filesystem truth. ``Drops'' are
accounting updates discarded by the never-blocking queue.}
\centering
\setlength{\tabcolsep}{3.5pt}
\scriptsize
\begin{tabular}{@{}L{2.1cm} c L{4.0cm}@{}}
\toprule
Pressure rule & Drops absorbed & Outcome \\
\midrule
index total only (na\"ive) & $\sim$\cfourDiedDrops &
partition 100\% full; snapshotter unstartable; fatal \texttt{ENOSPC} \\
+ drop estimate, raised to \texttt{statfs} when lossy; escalation rescan
& $\sim$\cfourHeldDrops &
\texttt{du} \cfourHeldDu{} vs. \cfourBudget{} budget; node startable;
rescans fired 3--4$\times$/arm; residual under-estimate
\cfourResidualLo--\cfourResidualHi{} on 3 of $\sim$95 samples \\
\bottomrule
\end{tabular}
\label{tab:cfour}
\end{table} The known residual is bounded and reported: pressure
under-estimated \texttt{du} by \cfourResidualLo--\cfourResidualHi{} on 3 of
$\sim$95 samples (chunks written during a rescan's own tree walk), never
approaching the partition headroom above the budget on this geometry. The complement also
held: with zero drops the filesystem is never consulted, so a shared
partition's foreign bytes cannot be charged to the cache.

\subsection{A second tenant, and a fleet}\label{sec:tenancy}
Two campaigns push past the single-pod, single-node envelope.

\textbf{Interference under a shared budget.} Pod A serves continuous
reads from a \thrashHotSetGB{} hot set while pod B sweeps the
140\,GB-class image through the shared 80\,GB budget (LRU, two
repetitions). The headline survives tenancy: \textbf{zero} read errors
on A across \intReads{} measured reads, while B completes 280/280
within 0.4\% of its single-tenant time. The interference is real,
modest, and replicates: A's p50 rose +\intPFiftyLo{} and +\intPFiftyHi{}
under the sweep (two repetitions agreeing to 0.3\,pp) and its cache hit
rate fell \intHitDropLo{} and \intHitDropHi{} points, attributed via
per-blob registry accounting (A's image shares only a 2.3\,MB base
layer with B's, tallied separately). A pre-specified p99 rule proved
under-powered at N=2---the baseline's own between-repetition p99 spread
exceeds most of the effect---and is reported as undecided rather than
leaned on; p50 and hit rate both replicate.

\textbf{The finding we did not go looking for.} In the baseline
phase---pod A alone on the node---its \thrashHotSetGB{} hot set occupied
$\sim$\thrashOccupiedGB{} of cache (\thrashFactor$\times$ its bytes),
pinning occupancy at the high watermark and evicting continuously with
no second tenant present (repetitions agree: 73.0 and 73.2\,GB,
$\sim$610{,}000 evictions per 300\,s window). A single read of the set
costs $2.0\times$---the known duplication across the HTTP-chunk and
filesystem cache tiers---and the remaining inflation accrues with
re-reads; this campaign did not isolate which tier dominates, or
whether the growth is stale copies or accounting. Effective capacity
for this workload shape is roughly budget/\thrashFactor. This reorders
I6's question: before asking whether a sweep may flush \emph{another}
tenant's hot set, a tenant whose hot set is 27\% of the nominal budget
already cannot keep it. Both interference numbers above are therefore
measured on top of an already-thrashing tenant, not against a quiescent
warm baseline; the pre-specified ``zero refetches in steady state''
entry condition was unsatisfiable on this geometry and is reported as
such rather than silently relaxed.

\textbf{Registry amplification at $k=3$.} Three nodes, each with its
own 80\,GB budget at 90\% pre-fill, read the full image simultaneously
(start skew $\leq$20\,ms) from one registry host. Every node's budget
held with zero errors (2{,}520 reads). Aggregate registry egress was
linear in $k$ to within \ampLinearityHi/\ampLinearityLo{} of $3\times$
the same-session $k{=}1$ baseline: lazy pulling under exhaustion
multiplies registry load with no cross-node relief, as it must with
node-local caches and no peer-to-peer layer. The registry's measured
10.3\,Gbps NIC reached only \ampNicPct{} at $k{=}3$, and the
pre-specified expectation that concurrent readers would slow each node
was falsified: all six comparisons ran
\ampFasterLo--\ampFasterHi{} \emph{faster} than the same node's own
$k{=}1$ run. We hypothesize registry-side page-cache warmth across
rounds and did not instrument it; the claims this campaign licenses are
the linearity, the held budgets, and the unsaturated NIC. The
saturation point lies beyond three nodes on this hardware and is not
extrapolated.

\subsection{Eviction policy: an honest negative}
The mechanism takes pluggable policies, and serving workloads are
scan-shaped, so we expected scan-resistant eviction (a 2Q-style
probationary/protected split: a chunk must be touched twice before it
earns protection from a passing sweep) to beat LRU. It did not. The
comparison is disciplined by a gate, pre-specified for the definitive
runs though instituted mid-investigation: before any between-policy
difference may be interpreted, every arm's warm baseline run---same
build, policy irrelevant---must agree within 5\%, since a difference
between identical configurations is the rig's noise floor speaking. The
gate exists because a first attempt produced an apparent 2Q win that was
traced to a baseline mismatch and withdrawn. Under the gate, eight
interleaved arm-runs (four per policy) shared a warm baseline to within
\policyGatePct, and under the pre-specified effect-size rule---a
between-policy difference counts only if it exceeds the spread between
repetitions of the same policy---LRU and 2Q were not separable at p50,
p95, p99, or hit rate. With the noise floor at \policyGatePct, this
campaign's minimum detectable effect is roughly 5\%: we claim
equivalence within that bound, not the absence of any effect. A
follow-up in a deliberately contended regime (70\,GB hot set against the
80\,GB budget, $\sim$10.5\,M evictions per arm) failed its own gate
(\policyGateFailPct) and is reported only as consistent with the null.
Across the entire campaign no measurement produced positive evidence for
2Q. We ship LRU as the default and report the policy interface as
mechanism, not advantage.

\subsection{Overheads}\label{sec:overheads}
Accounting costs +\mbGetNs\,ns (+\mbGetPct) per cache-hit read and
+\mbAddUs\,\textmu s (+\mbAddPct) per cache write in microbenchmarks; the
filesystem-truth pressure path added \abGapPct{} in an interleaved
same-session A/B at full sweep scale. A 30\,s race-detector stress run
served \stressReads{} reads while \stressEvictions{} chunks were evicted
underneath with zero incorrect bytes and zero non-miss errors.

\section{Discussion: What Cloud Platforms Should Take}\label{sec:discussion}
Three lessons generalize beyond \stargzsn, and one objection deserves a
direct answer.

\textbf{Align the actuator with the accounting, or own the bytes.} F4 is an
instance of a general hazard: a platform whose reclamation actuator
(kubelet image GC, pod eviction) is keyed to one accounting (containerd
images, pod ephemeral storage) cannot manage a resource that lives outside
it, no matter how good its signal is. Wherever a node component maintains
a shared on-disk cache---snapshotters, CSI drivers, model caches---either
its bytes must appear in an accounting the platform can act on, or the
component must bound itself. We chose the second because it needs no
kubelet or CRI changes; the first is the longer-term fix and worth a
platform-level design discussion.

\textbf{Cache failure must be decoupled from data failure by construction,
not by monitoring.} F3 shows the detection route is sampling-limited even
when built in good faith; the pass-through result shows the construction
route is cheap (+\honestyPricePct{} at total exhaustion, \abGapPct{} when
healthy). We suggest this as a design test for any read-through cache on a
critical data path: inject a full cache volume and ask whether reads fail.
The audit of Section~\ref{sec:audit} grades the ecosystem against it:
Nydus RAFS passes by construction, stargz and SOCI fail, and fscache mode
cannot be graded from user space.

\textbf{Truth for \emph{how much} and truth for \emph{which} can come from
different sources.} An accounting that must never block the data path will
be lossy; an enforcement decision that trusts it inherits the loss at the
worst moment. Splitting the roles---filesystem-level usage decides when to
evict, the index decides what---kept the budget standing under
$\sim$\cfourHeldDrops{} dropped updates. The same split should apply to
any advisory index guarding a hard resource.

\textbf{Sidestepping the cache does not dissolve the lesson.} An
operator can avoid this failure class by not routing weights through the
image at all---volume-based model storage, or the OCI image volume
source. Those designs trade away what lazy pulling is for
(registry-native distribution, provenance, and the cold-start result
that motivates this line of work~\cite{kliukovkin2026coldstart}) and
move the same question elsewhere: whatever component then caches model
bytes on the node faces the same bound-and-honesty test. Likewise the
folk remedy---a periodic prune of the cache tree---is an unsynchronized
actuator with no accounting: it cannot respect in-flight pins or
recency, deletes at arbitrary points in the write path, and bounds
nothing between prunes. We did not evaluate it, but nothing in it
addresses the write-failure coupling that produces F2.

\section{Related Work}\label{sec:related}
\textbf{Lazy image distribution.} Slacker~\cite{slacker} established lazy
container distribution; DADI~\cite{dadi}, Starlight~\cite{starlight},
FaaSNet~\cite{faasnet}, Nydus~\cite{nydus}, \estargz~\cite{estargz}, and
SOCI~\cite{soci} developed formats and fetching architectures. That
literature evaluates start-up acceleration; our companion
study~\cite{kliukovkin2026lazy} measured its deferred cost and failure
semantics, and this paper closes the loop with a fix. \textbf{Bounded
caches for remote filesystems.} CernVM-FS has long shipped an LRU-evicted,
quota-bounded local cache for a FUSE-mounted remote
filesystem~\cite{cvmfs}; we adopt the spirit and differ in what forced the
design: a Kubernetes node's cache is multi-tenant across pods, its
accounting must never block a serving data path (hence lossy accounting
and filesystem-truth pressure, Section~\ref{sec:c4}), and its failure mode
couples into pod lifecycle semantics rather than a single user's job.
\textbf{Cache replacement.} The policy interface implements LRU and a
2Q-style split~\cite{twoq}; scan-resistant designs from ARC~\cite{arc}
to the recent S3-FIFO~\cite{s3fifo} and SIEVE~\cite{sieve} target
exactly the access patterns where LRU is weakest. Our negative result
does not contradict that literature: on a serving node dominated by
large sequential sweeps through a budget smaller than the working set,
the gains scan resistance buys on skewed web workloads have nothing to
grip---no policy can rescue a sweep that must evict its own tail (I6).
\textbf{Mount survival across daemon restarts} is prior art we adopt, not
claim: Nydus's failover~\cite{nydus} and JuiceFS's smooth
upgrade~\cite{juicefs} both hand off live FUSE sessions (the FUSE
data path's costs are themselves well characterized~\cite{fuse17});
\stargzsn's own FUSE manager does the
same---F5's finding is that its shipped packaging silently defeats it, and
its graceful-restart cache duplication was undocumented. \textbf{Kubernetes
resource management and failure semantics.} Node-pressure eviction and
image GC~\cite{kubelet}
manage disk the platform can see; F4 measures precisely what happens when
they meet a resource they cannot reclaim. The green-while-failing pods of
F2 and kubelet's unreclaiming eviction loop of F4 are instances of
patterns documented at scale---fail-slow rather than fail-stop
behavior~\cite{failslow} and sustained metastable
degradation~\cite{metastable}; our contribution is tracing one such loop
to an actuator/accounting mismatch and removing the trigger by
construction. Work on LLM serving cold starts
(e.g., ServerlessLLM~\cite{serverlessllm}) attacks load-time from the
checkpoint side and is complementary: our concern is what the delivery
cache does to a fleet after \texttt{Ready}. \textbf{Management of
lazy pulling and Kubernetes fleets.} Chen \emph{et al.}~\cite{chen2024lazy}
manage lazy pulling across edge nodes by sharing image caches
collaboratively to shorten start-up; KubeTwin~\cite{kubetwin} models
Kubernetes deployments as digital twins to predict their behavior at
scale. Both treat the node-level cache as a resource to be placed and
sized; neither asks what the platform observes, or can do, once that
cache exhausts its volume, which is the management gap this paper
measures and closes.

\section{Limitations}\label{sec:limits}
The failure mechanism is node-local, and the multi-node campaign
measured three same-AZ nodes: registry saturation ($k^{*}$), cross-AZ
registries, and placement consequences of budget pressure (a scheduler
declining to place lazy pods on saturated caches) remain unevaluated,
and the $k{=}3$ speed-up's suspected registry page-cache cause is
uninstrumented. The two-tenant campaign measured interference on top of
an already-thrashing tenant (Section~\ref{sec:tenancy}); its p99 rule
was under-powered at N=2, and the \thrashFactor$\times$ cache inflation
behind the thrash---which tier dominates, stale copies versus
accounting---is the paper's most important open measurement.
One hardware generation (i4i NVMe); absolute
times scale with the rig, which is why only within-session comparisons are
quoted, and the +\honestyPricePct{} price of exhaustion is specific to a
same-AZ registry. The elimination result is established on the same
failure suite that documented the problem; the claim that the class is
closed for other workload shapes rests on the mechanism argument of
Section~\ref{sec:rootcause}---the severed coupling is the only path by
which a cache write error reached the reader---together with the
\texttt{ENOSPC} regression test, not on workload breadth. The
write-flood result bounds the budget's survival at
$\sim$\cfourHeldDrops{} dropped updates but does not establish the rate at
which reconciliation stops converging; that ceiling is unmeasured. The
policy negative is established on one gate-passing workload family
(four arm-runs per policy) with a contended follow-up that failed its own
gate; it licenses ``no observed advantage exceeding the $\sim$5\%
minimum detectable effect,'' not ``no possible advantage.'' The
\cfourResidualLo--\cfourResidualHi{} reconciliation residual was measured
on one partition geometry with 12\,GB of partition headroom above the
budget; tighter geometries deserve their own check. The cross-snapshotter audit is
static: SOCI's and Nydus's behavior under runtime exhaustion---including
whether SOCI shares the sampling-blindness of F2's damage
pattern---was not measured here, and four specific questions the sources
cannot answer are flagged as such in the released audit document; the
managed lazy-pulling offerings of cloud providers (GKE Image Streaming,
Azure Artifact Streaming, SOCI on EKS and Fargate) are closed-source
beyond the SOCI component and could not be audited. The budget is a
single per-node bound: per-tenant or per-namespace quotas, fairness, and
noisy-neighbor isolation are not provided, and the
\thrashFactor$\times$ inflation finding makes them the natural next step
for the policy interface. The $k{=}3$ simultaneous exhausted reads are
the closest this evaluation comes to a scale-out burst of concurrent
cold starts under budget pressure; an explicit autoscaling campaign
remains to be run. Cost modelling---translating the measured
+\honestyPricePct{} completion cost and linear registry egress into
provider prices---is out of scope; the quantities needed for it are
reported.

\section{Conclusion}\label{sec:concl}
Lazy pulling's cold-start win for model serving was never in doubt; its
safety was. We traced the silent corruption of already-serving pods to a
single coupling on the fetch path, severed it with pass-through semantics,
and bounded the cache with an eviction engine that survives its own lossy
accounting by trusting the filesystem for how much and the index for
which. On the failure suite that documented the problem, the failure class
is gone: \oursOkFiles/280 files served across a \oursBytesRead{} read at
90\% pre-fill, against \vanillaOkFiles/280 upstream, for a
+\honestyPricePct{} completion cost and data-path overheads of
+\mbGetNs\,ns per read and +\mbAddUs\,\textmu s per write. The platform above cannot do this on
the cache's behalf---given perfect visibility it freed nothing by image
garbage collection and its pod evictions released no cache bytes---so the
bound belongs where the bytes live. Harness, raw data,
evidence bundles, and the implementation are open source at
\url{https://github.com/kliukovkin/lazy-model-delivery-study} and the
\stargzsn fork referenced there.

\section*{Acknowledgment}
The author discloses that AI-based assistants (Anthropic Claude) were used
in developing the implementation, benchmark tooling, and analysis scripts
and in editing the manuscript text; no section of this article was
generated by an AI system. The study design, the experiments, validation
of all results, and responsibility for the content rest with the author.
All raw data, per-trial evidence bundles, pre-specified protocols, and the
measurement harness are publicly available at
\url{https://github.com/kliukovkin/lazy-model-delivery-study} (paper3/).
The author declares no competing interests and received no funding for
this work.

\begin{IEEEbiographynophoto}{Georgii Kliukovkin}
is an independent researcher in Sunnyvale, CA, USA. His work concerns
container image distribution, model-serving infrastructure on Kubernetes,
and the dependability of cloud-native systems, and includes merged
contributions to the KServe project (OCI-based model storage) and
contributions proposed to containerd/stargz-snapshotter. He holds a
master's degree.
\end{IEEEbiographynophoto}


\begin{thebibliography}{00}
\bibitem{kliukovkin2026coldstart} G. Kliukovkin, ``Cold-start model delivery
in Kubernetes inference serving: an empirical study of OCI-based distribution
and its integrity,'' \emph{IEEE Access}, vol. 14, 2026, doi:
10.1109/ACCESS.2026.3731156.
\bibitem{kliukovkin2026lazy} G. Kliukovkin, ``The lazy pod that lies:
deferred cost and failure semantics of lazy container image pulling for
model serving on Kubernetes,'' 2026, arXiv:2608.19412.
\bibitem{estargz} containerd project, ``stargz-snapshotter: fast container
image distribution plugin with lazy pulling,''
\url{https://github.com/containerd/stargz-snapshotter}, v0.18.2, 2026.
\bibitem{soci} AWS Labs, ``soci-snapshotter: a containerd snapshotter plugin
for lazy loading with the Seekable OCI format,''
\url{https://github.com/awslabs/soci-snapshotter}, 2026.
\bibitem{slacker} T. Harter, B. Salmon, R. Liu, A. C. Arpaci-Dusseau, and
R. H. Arpaci-Dusseau, ``Slacker: fast distribution with lazy Docker
containers,'' in \emph{Proc. 14th USENIX FAST}, 2016, pp. 181--195.
\bibitem{dadi} H. Li \emph{et al.}, ``DADI: block-level image service for
agile and elastic application deployment,'' in \emph{Proc. USENIX ATC},
2020.
\bibitem{starlight} J. L. Chen, D. Liaqat, M. Gabel, and E. de Lara,
``Starlight: fast container provisioning on the edge and over the WAN,'' in
\emph{Proc. USENIX NSDI}, 2022, pp. 35--50.
\bibitem{faasnet} A. Wang \emph{et al.}, ``FaaSNet: scalable and fast
provisioning of custom serverless container runtimes at Alibaba Cloud
Function Compute,'' in \emph{Proc. USENIX ATC}, 2021.
\bibitem{nydus} Dragonfly project, ``Nydus: the Dragonfly image service,''
\url{https://github.com/dragonflyoss/nydus}, 2026.
\bibitem{cvmfs} J. Blomer, P. Buncic, and T. Fuhrmann, ``CernVM-FS:
delivering scientific software to globally distributed computing
resources,'' in \emph{Proc. 1st Int. Workshop Netw.-Aware Data Manage.
(NDM)}, 2011, pp. 49--56.
\bibitem{kubelet} Kubernetes Documentation, ``Node-pressure eviction,''
\url{https://kubernetes.io/docs/concepts/scheduling-eviction/node-pressure-eviction/},
2026.
\bibitem{serverlessllm} Y. Fu \emph{et al.}, ``ServerlessLLM: low-latency
serverless inference for large language models,'' in \emph{Proc. USENIX
OSDI}, 2024, pp. 135--153.
\bibitem{twoq} T. Johnson and D. Shasha, ``2Q: a low overhead high
performance buffer management replacement algorithm,'' in \emph{Proc.
20th Int. Conf. Very Large Data Bases (VLDB)}, 1994, pp. 439--450.
\bibitem{arc} N. Megiddo and D. S. Modha, ``ARC: a self-tuning, low
overhead replacement cache,'' in \emph{Proc. 2nd USENIX FAST}, 2003,
pp. 115--130.
\bibitem{s3fifo} J. Yang, Y. Zhang, Z. Qiu, Y. Yue, and R. Vinayak,
``FIFO queues are all you need for cache eviction,'' in \emph{Proc. 29th
ACM SOSP}, 2023, pp. 130--149.
\bibitem{sieve} Y. Zhang, J. Yang, Y. Yue, Y. Vigfusson, and
K. V. Rashmi, ``SIEVE is simpler than LRU: an efficient turn-key
eviction algorithm for web caches,'' in \emph{Proc. 21st USENIX NSDI},
2024, pp. 1229--1246.
\bibitem{fuse17} B. K. R. Vangoor, V. Tarasov, and E. Zadok, ``To FUSE
or not to FUSE: performance of user-space file systems,'' in \emph{Proc.
15th USENIX FAST}, 2017, pp. 59--72.
\bibitem{failslow} H. S. Gunawi \emph{et al.}, ``Fail-slow at scale:
evidence of hardware performance faults in large production systems,''
in \emph{Proc. 16th USENIX FAST}, 2018, pp. 1--14.
\bibitem{metastable} N. Bronson, A. Aghayev, A. Charapko, and T. Zhu,
``Metastable failures in distributed systems,'' in \emph{Proc. 18th
Workshop Hot Topics Oper. Syst. (HotOS)}, 2021, pp. 221--227.
\bibitem{juicefs} Juicedata, ``JuiceFS: a distributed POSIX file system
built on top of object storage,''
\url{https://github.com/juicedata/juicefs}, 2026.
\bibitem{chen2024lazy} C.-C. Chen, Y. Chiang, Y.-C. Lee, and H.-Y. Wei,
``Edge computing management with collaborative lazy pulling for
accelerated container startup,'' \emph{IEEE Trans. Netw. Service
Manag.}, vol. 21, no. 6, pp. 6437--6450, Dec. 2024.
\bibitem{kubetwin} D. Borsatti \emph{et al.}, ``KubeTwin: a digital twin
framework for Kubernetes deployments at scale,'' \emph{IEEE Trans. Netw.
Service Manag.}, vol. 21, no. 4, pp. 3889--3903, Aug. 2024.
\end{thebibliography}
\end{document}